\documentclass[acmsmall]{acmart}

\usepackage{subcaption}
\AtBeginDocument{%
  \providecommand\BibTeX{{%
    \normalfont B\kern-0.5em{\scshape i\kern-0.25em b}\kern-0.8em\TeX}}}

\setcopyright{none}

\renewcommand\footnotetextcopyrightpermission[1]{} 
\begin{document}


\title{Em-K Indexing for Approximate Query Matching in Large-scale ER}


\author{Samudra Herath}
\affiliation{%
  \institution{University of Adelaide}
  \streetaddress{North Terrace}
  \city{Adelaide}
  \state{South Australia}
  \country{Australia}
  \postcode{5000}
}
\email{samudra.herath@adelaide.edu.au}
\orcid{1234-5678-9012}

\author{Matthew Roughan}
\affiliation{%
  \institution{University of Adelaide}
  \streetaddress{North Terrace}
  \city{Adelaide}
  \state{South Australia}
  \country{Australia}
  \postcode{5000}
}
\email{matthew.roughan@adelaide.edu.au}

\author{Gary Glonek}
\affiliation{%
  \institution{University of Adelaide}
  \streetaddress{North Terrace}
  \city{Adelaide}
  \state{South Australia}
  \country{Australia}
  \postcode{5000}
}
\email{gary.glonek@adelaide.edu.au}

\renewcommand{\shortauthors}{Herath, et al.}

\begin{abstract}

Accurate and efficient entity resolution (ER) is a significant challenge in many data mining and analysis projects requiring integrating and processing massive data collections. It is becoming increasingly important in real-world applications to develop ER solutions that produce prompt responses for entity queries on large-scale databases. Some of these applications demand entity query matching against large-scale reference databases within a short time. We define this as the query matching problem in ER in this work. Indexing or blocking techniques reduce the search space and execution time in the ER process. However, approximate indexing techniques that scale to very large-scale datasets remain open to research. In this paper, we investigate the query matching problem in ER to propose an indexing method suitable for approximate and efficient query matching.

We first use spatial mappings to embed records in a  multidimensional Euclidean space that preserves the domain-specific similarity. Among the various mapping techniques, we choose multidimensional scaling. Then using a Kd-tree and the nearest neighbour search, the method returns a block of records that includes potential matches for a query. Our method can process queries against a large-scale dataset using only a fraction of the data $L$ (given the dataset size is $N$), with a $O(L^2)$ complexity where $L \ll N$. The experiments conducted on several datasets showed the effectiveness of the proposed method.




\end{abstract}



\maketitle 

\section{Introduction}

Many real-world data collections are low in quality because of errors (e.g., typographical errors or phonetic misspellings), incomplete or missing data, incompatible formats for recording database fields (e.g., dates, addresses), and temporal inconsistencies. Integrating or querying multiple data sources to identify records that belong to the same real-world entity is a challenging task in the presence of such data. This task is often referred to as the \textit{entity resolution (ER)} problem and appears in many database applications when identifying duplicates, cleansing data, or improving data quality.

ER has been studied for years. However, the issues of efficiently handling large-scale data remain an open research problem~\cite{Liang}. Typical ER algorithms have a quadratic running time, which is computationally prohibitive for large-scale data collections. This performance bottleneck occurs due to the detail-level pairwise comparison step of the ER process.

Consider the motivating example of linking medical records where one entity can have many entries over several years. A public health case study could involve hundreds of thousands of entities with millions of records. Quadratic ER among the records will require trillions of pairwise comparisons in this application, which is computationally infeasible.

Indexing techniques address this problem by grouping similar records into blocks. Usually, groups of similar records within or between datasets are smaller than the total number of records. Hence, many comparisons will be among the non-matching records when applying pairwise comparisons using a brute-force ER algorithm. Indexing techniques aim to reduce the potential number of comparisons by reducing the comparisons between those non-matches~\cite{sur1}. Therefore, indexing-based ER solutions run much faster than brute-force ER solutions~\cite{HARRA}.

However, most existing indexing methods still have quadratic time complexity and are too slow to deal with very large-scale data collections. Also, some non-scaling ER algorithms tend to generate a large number of blocks and a large number of candidate records in each block for large-scale scenarios~\cite{sur1}.

Traditional ER solutions often process databases offline in batch mode, and no further action is required once a pair of matches are determined. However, many organisations are moving online where they have to provide their services through prompt responses. Hence, many newer, real-world scenarios require real-time query processing against large-scale databases. Some of the applications demand entity query matching against large-scale reference databases within a short time. We refer to this as the \textit{query matching} problem in ER. In such configurations, traditional ER solutions become less efficient or unusable. 


In order to motivate the problem context and illustrate the usefulness of the approach presented in this paper, we provide the following real-world example. The application is in criminal investigations, where law enforcement officers need to query a database to identify potential suspects. Suspects usually lie to police investigators about their true identities, e.g., names, birth dates, or addresses~\cite{Chen_Chung}. Hence, finding an exact match for falsified data against the real identity recorded in a law enforcement computer system is problematic. Detecting deceptive identities is a time-consuming activity that involves large amounts of manual information processing in real-life scenarios~\cite{cr_1}. Therefore efficient ER solutions that support real-time and approximate query matching against existing datasets can be invaluable.


Toward the challenge of real-time approximate query matching, we present an indexing technique that reduces the number of pairwise comparisons needed in the ER. The proposed indexing method transforms a set of records into a set of vectors in a metric-space, specifically a lower-dimensional Euclidean space using multidimensional scaling. These vectors have two main attractive attributes in the context. First, comparisons between vectors in a metric-space are much cheaper than string comparisons. Second, these vectors support efficient indexing data structures. We utilise these properties to propose an indexing approach that classifies similar vectors in a low-dimensional Euclidean space. We call our method \textit{Em-K indexing}, as our method operates by embedding data in a K-dimensional space.

Exact query matching is easy against a reference database as we can search based on lexicographical order using an efficient data structure, e.g., binary trees. However, exact query matching is impossible with many real-world databases due to various data quality issues, and approximate query matching is the common approach~\cite{Christen2012}. The proposed indexing method address the problem of real-time and approximate query matching. We will be mainly using the term query matching instead of approximate query matching in this paper.

Our main objective is to develop a fast indexing method for query matching. We want our method to be efficient for large-scale data and robust to errors in the data. The proposed method searches for a block of records in the reference database as a set of potential matches to the querying record. Hence we avoid comparing the query against all the records in the referencing database. However, the search is done in the Euclidean space using a Kd-tree data structure where each record in the reference database and a query record become a vector in this space. 

Given the block size is~$B$, we search for this number of nearest neighbours for the query point by traversing the Kd-tree nodes that store the reference database. The search has a $O(B \log N)$ complexity, given $N$ is the number of records in the reference database. The method embeds query records in a pre-mapped Euclidean space to search against the reference data points. We propose an out-of-sample embedding that uses a fraction of the original data (defined as landmarks in Section~5) for query record embedding. For $L$ landmarks, the embedding requires $O(L^2)$ operations, and we can choose $L$ such that  $L \ll N$.

In this paper, we explore the use of metric-space indexing for efficient and approximate query matching. In particular, our contributions are,

\begin{itemize}

    \item We formulate the query matching and deduplication problems in ER to provide metric-space base solutions. First, we propose an indexing approach based on landmarks as a motivating example to explore the basic building blocks needed for query matching. 

    \item We propose a landmarks based indexing technique for query matching to provide a quick and accurate block of potential matches. Our method can process a stream of queries against a large-scale data set within a short time. By doing so, we obtain as many of the matching records as possible where the processing time of a single query takes a sub-second time. The technique is robust to noisy data that contains errors and allow efficient approximate query matching. 
    
\end{itemize}


\section{Preliminaries}

To describe the problem succinctly, we first describe relevant definitions and some key concepts. The definitions of relevant concepts here follow those in~\cite{schema_agnostic,sur2}. 

\subsection{Entity Resolution }
 
An {\it entity} is a real-world object, e.g., person, place or product, that has a unique identifier to distinguish it from other entities of the same type. An {\it entity profile} describes an entity using a collection of name-value pairs. A set of entity profiles is called {\it entity collection}, denoted by $E$. A pair of similar entity profiles are called {\it duplicates}. A duplicate of an entity can be either an exact copy of the original entity profile or an entity profile that contains an error (e.g., typographical error). A database representation of an entity profile is usually referred to as a \emph{record}. 

\newtheorem{mydef1}{Definition}
\begin{mydef1}(Entity Resolution): Given two records $r_i, r_j$ is a {\it match},
if they refer to the same unique real-world object. We denote this as
$r_i \equiv r_j$. The goal of ER is to link different records that describe the same entity within an entity collection or across two or more entity collections. 
\end{mydef1}


\subsection{ER Tasks}

Following the above definitions, we distinguish between the following ER tasks.

\begin{enumerate}

    \item \textit{Dirty ER:} Given an entity collection $E$ that contains entities $e_1,e_2,..., e_n$, find all duplicates and produces a set of equivalence clusters of distinct entities. It is also known as deduplication in many database applications.  
    
    \item  \textit{Clean–Clean ER:} Given two duplicate free entity collection $E_1$ and $E_2$ that contains entities $e_1,e_2,..., e_n$, find all records that belong to a single entity. Our indexing method is aimed at the \textit{Clean–Clean ER} problem where one dataset is a stream of queries, and the other is a reference database. However, we use deduplication as a motivating example that explains the relevant building blocks of the proposed method. 
    
\end{enumerate}

\subsection{Record Comparisons} 

Similar records pairs of an entity are determined by applying a similarity function over the corresponding attributes of two records. Assume a pair of records  $(r_i, r_j)$ and a set of attributes $a_1, a_2, ... , a_x$ that describe them. The similarities $s_1, ..., s_x$ between attribute values are determined by applying a set of similarity functions $sim(r_i.a_k , r_j.a_k )$, with $1 \leq k \leq x$ for each pair of attribute values. Then a total similarity score of $S=\sum_{i=1}^{x}s_i$ is calculated to classify the record pairs as a match or a non-match based on a matching threshold.

Several comparison methods such as edit distance, a.k.a Levenshtein distance, Jaro distance, and q-gram distance are found in the domain of strings~\cite{Loo2014}. In this work, we mainly used Levenshtein distance to measure the similarity at the attribute level. It calculates the minimum number of character insertions, deletions, and replacements necessary to transform a string $s_1$ into a string $s_2$. Minkowski metrics based on $L^p$ norms, ${\parallel x \parallel}_p = (\sum_i |x_i|^p )^{1/p}$, with $p \geqslant 1$ are a common used class of vector spaces. For our vector space, we used the most common Minkowski metric; Euclidean distances $d_E(p=2)$.


\subsection{Indexing or Blocking Techniques:}

Traditional ER requires pairwise comparisons between all the records. For instance, gievn two entity collections $E_1$ and $E_2$, with sizes $|E_1|$ and $|E_2|$, it requires $|E_1| \times |E_2|$ comparisons between entity records. In practice, this is infeasible when, $|E_1|$ and $|E_2|$ are large due to the inherent quadratic complexity of the comparison process.

Indexing or blocking reduces the number of detail-level pairwise comparisons between records by removing pairs that are unlikely to be real matches. The traditional blocking techniques partition the databases into non-overlapping blocks, only comparing the records within blocks. Hence, reducing the number of pairwise comparisons.

Our method is an example of join-based blocking techniques that convert blocking into the nearest neighbour search. The blocks are created by searching the vector space for similar records using the Em-K indexing rather than partitioning the dataset. As a result, similar records are grouped into overlapping blocks by combining spatial joins with block building~\cite{sur2}.  



\section{Problem Formulation}

In this section, we formulate our problem using the concepts and the definitions presented in the previous section. The proposed Em-K indexing method functions as a preprocessing approach for a more detailed query-matching ER. We use deduplication as a motivating example that explains the basic building blocks to the query matching problem. Thus we defined the following two problems:

\medskip
\noindent \textbf{Query Matching}: is the problem of finding similar records given two entity collections, $E_r$ a reference database, and $Q$ a stream of queries. The size of each dataset is denoted by $|E_r|$ and $|Q|$ respectively and we assume $|E_r|$ is fixed and $|Q|\to \infty$. Let $Q={q_1,q_2,..,q_{|Q|}}$, where each query $q_i$ in $Q$ represents a record of an entity $e_j$. A query record $q_i$ has the same attribute schema as the records in $E_r$. Hence for each query $q_i$ in $Q$, the records in $E_r$ that belong to the same entity ($e_j$) need to be found. In real-life problems, the query rate might be very fast. There may or may not be a matching record for every $q_i \in Q$ that belongs to the entity $e_j$.

\newtheorem{mydef}{Problem Statement}
\begin{mydef}
(Indexing for Query matching): For two datasets (one is a reference database and the other is a stream of queries) with overlapping records, run the best algorithm for a given amount of time to find the block of records containing as many matches as possible for the querying record. This is a pre-processing step to increase the efficiency of the subsequent detailed query match. 
\end{mydef}



We also consider the deduplication problem here, primarily as an explanatory model. We defined deduplication in Section 2.2 under \textit{dirty ER}. In the following, we define the indexing for deduplication.

\begin{mydef}
(Indexing for deduplication): For a given entity collection containing duplicates, group similar records into blocks to reduce the number of comparisons needed in subsequent detailed deduplication while missing as few matches as possible.
\end{mydef}

Traditional indexing techniques split the database into non-overlapping blocks, only comparing the records within any block~\cite{MacCallum, sur1, sur2, Matthew, Naumann}. Blocks of similar pairs are determined by building an indexing structure that takes a set of records as input and classifies them according to some criteria. Usually, this criterion is based on matching a \textit{blocking key} consisting of a single or several attribute values of records~\cite{sur1}. Our method uses blocking values to create blocks of records that transform blocking into a k-nearest neighbour (k-NN) search. We combine the spatial joins with block building to convert blocking values to a similarity preserving Euclidean space. The result is overlapping blocks of records.

The proposed method requires mapping blocking values into multidimensional vectors. Since many comparisons in ER are between strings of characters, we focus on entity attributes that contain string values here. Unlike the similarity between string values, the similarity between numerical values is easy to compute using $L^p$ norms in a metric-space~\cite{HARRA}. Hence, we used this property of the metric-space to propose a scalable indexing technique for query matching.

If we assume strings as elements of a complicated high-dimensional space, the distance between two different strings is typically large. However, misspelled strings tend to be located near correctly spelled strings~\cite{Mazeika}. The embedding of a string database into a metric-space needs to preserve these two properties. Thus, coordinates for blocking values are determined in a Euclidean space such that the associated Euclidean distances approximate the dissimilarities between the original blocking values.  



The general problem of assigning coordinates in this manner is one of embedding a metric or non-metric-space into a Euclidean space~\cite{Virtual_L}. Suppose $R$ is a collection of objects, $\delta$ measures the distances between $R$ objects, $X$ represents the coordinates matrix for the $R$ objects in the Euclidean space, and $d$ measures the distances between coordinates.

Embedding of a metric or non-metric-space $(R,\delta)$ into a Euclidean space $(X,d)$ is a mapping $\phi : R \to X$. In this paper $(R,\delta)$  will always be a finite space (\textit{i.e.,} $R$ is a finite set) and $(X,d)$ will always be a Euclidean space.

\begin{mydef}
(The embedding problem): For a given metric or non-metric space, find a $\phi$ that minimizes distortion, stress, or a similar error metric between $(R,\delta)$ and $(X,d)$.
\end{mydef}

A commonly used technique for embedding a set of distances (or dissimilarities) into a Euclidean space is multidimensional scaling (MDS)~\cite{Kruskal1964}. We applied MDS because we can adapt it to achieve good time efficiency and distance preserve capability for large-scale data with a small amount of extra effort.

Among the variants of MDS, we use least-squares multidimensional scaling (LSMDS) for the embedding since it gave the best results compared to other variations such as classical scaling~\cite{herath2020simulating}. We can map a set of blocking values to a lower-dimensional Euclidean space by applying LSMDS such that the distances between vectors preserve the dissimilarities between them. This embedding leaves similar blocking values closer in the Euclidean space allowing efficient, geometric-based indexing.

LSMDS initially maps each item in the non-metric or metric-space to a $K$-dimensional point. Then minimises the discrepancy between the actual dissimilarities and the estimated distances in the $K$-dimensional space by optimisation~\cite{MDS}. This discrepancy is measured using \textit{raw stress}~($\sigma_{raw}$) given by the relative error where ${\delta_{ij}}$ is the dissimilarity between the two objects and ${d_{ij}}$ is the Euclidean distance between their estimated points. 

\begin{equation}  \label{eq:1}
\sigma_{raw}(\mathbf{X}) ={\sum_{i,j=1}^{n}w_{ij}\Big(d_{ij}(\mathbf{X})-\delta_{ij}\Big)^{2}}.
\end{equation}

Possible weights for each pair of points are denoted by $w_{ij}$. Weights are useful in handling missing values and the default values are \mbox{$w_{ij} = 0$}, if $\delta_{ij}$ is missing and \mbox{$w_{ij} = 1$}, otherwise~\cite{MDS}. We do not apply weights in this work, hence, \mbox{$w_{ij} = 1$} always. We prefer the normalized stress ($\sigma$) in our experiments since it is popular and theoretically justified. The normalized stress ($\sigma$) is obtained by $\sigma= \sqrt{\sigma_{raw}(\mathbf{X})/\delta_{ij}^2 }.$

However, traditional MDS algorithms such as LSMDS require extensive preprocessing and usually are computationally expensive, thus not appropriate for large scale applications. The two main drawbacks are,

\begin{itemize}
    \item MDS requires $O(N^2 )$ time, where $N$ is the number of items. Thus, it is impractical for large~$N$.
    \item In an out-of-sample setting or a query-by-example setting, a query item has to be mapped to a point in the pre-mapped Euclidean space. Given LSMDS algorithm is $O(N^2)$, an incremental algorithm to search/add a new item in the database would be $O(N)$. Hence a query search would be similar to sequential scanning of a database~\cite{FastMap}.
\end{itemize}

Among the proposed methods of scalable MDS, we are interested in using an out-of-sample embedding approach as a scaling method for LSMDS. We have two main purposes:
\begin{enumerate}
    \item To embed large-scale reference databases.
    \item To embed previously unseen data to a pre-mapped Euclidean space. 
\end{enumerate}

Suppose we have a configuration of $N$ points in a \mbox{$K$-dimensional} Euclidean space obtained by applying LSMDS to a set of $N$ objects. Let $Q$ be out-of-sample objects, with measured pairwise dissimilarities from each of the original $N$ objects. The out-of-sample embedding problem is to embed the new $Q$ objects into the pre-mapped \mbox{$K$-dimensional} Euclidean space.

Our out-of-sample embedding approach uses the stochastic gradient descent algorithm to minimise the following objective function for numerical optimisation. The out-of-sample embedding of a new object $y$ is obtained by minimising the following objective function,

\begin{equation} \label{eq:2}
    \hat{\sigma}(y) = {\sum_{i=1}^n}\big( \left\| x_i-\hat{y}\right\|_2-\delta_{iy}\big)^2,
\end{equation}

\noindent where $y$ is the new object and $\hat{y}$ is its position in the Euclidean space. The $\delta_{iy}$ represent the dissimilarities between point $i$ and the new object $y$. The Euclidean distance between the $i^{th}$ point and $y$ in the Euclidean space is given  by $\left\| x_i-\hat{y}\right\|_2$. We seek to find a position of $y$ that minimises $\hat{\sigma}(y)$. Here we keep $w_{ij}=1$, similar to LSMDS.

The out-of-sample embedding approach becomes inefficient for large-scale data by comparing each new point with all the existing points. We scale our out-of-sample embedding solution to accommodate large-scale data by only considering a fraction of the pre-mapped Euclidean space. The initially selected subset of pre-mapped data is usually known as \emph{landmarks}. Landmarks or \emph{anchors} have been used with out-of-sample extensions to scale MDS and other embedding techniques~\cite{Silva_2002,Virtual_L}. We discuss the characteristics of a good set of landmarks and landmarks selection methods in Section~5.

Once the Euclidean space consists of all the data required, we then formulate our indexing method to generate blocks that group close-by vectors in the Euclidean space.

MDS maps high dimensional data (in the original space) into a Euclidean space (vector space). The rationale for performing such a mapping is to approximate the distances between objects in an original space in Euclidean space. Searching for similar points in Euclidean space is less expensive and quick since we can use efficient data structures such as Kd-trees. We use Kd-trees and k-NN search to find similar vectors in the Euclidean space.

In general, k-NN search refers to finding the closest elements for a query $q$ within a given set of points $N$, as measured by some distance function ${d}(N, q)$~\cite{Kumar}. The distance function $d$ is a metric, e.g., $L^p$ norm, which satisfies the non-negativity, identity, symmetry, and triangle inequality properties~\cite{JChatfied1980, Loo2014}. We use Euclidean distances $d_E$( where $p=2$) in our calculations. Here we are interested in finding the k-NNs ($k$ nearest neighbours) where k may be moderately large.


\begin{mydef}{k-nearest neighbour (k-NN) search}: The query retrieves the $k$ closest elements to $q$ in $N$. If the collection to be searched consists of fewer than $k$ objects, the query returns the whole database. If the set of the $k$ nearest neighbours of $q$ are $n_c$, then formally, $n_c$ can be defined as follows: 
\\
$kNN(q) = {n_c \subseteq N, |n_c| = k \land \forall x \in n_c, y \in N - n_c : d(q, x) \leq d(q, y)}$. 
\end{mydef}

An index is a data structure that reduces the number of distance evaluations needed at query time. An efficient and scalable indexing method can facilitate accurate and efficient k-NN search that supports large-scale datasets. We can apply several k-NN search methods for indexing arbitrary metric spaces; for more details, refer to the surveys~\cite{Hjaltason,Edgar}. Distance-based indexing methods use distance computations to build the index. Once the index is created, these can often perform similarity queries with a significantly lower number of distance computations than a sequential scan of the entire dataset~\cite{Hjaltason}.

The decision of which indexing structure to apply depends on several factors, including query type, data type, complexity and the application. Among many data structures, we choose Kd-trees for the k-NN search. It is considered one of the best data structures for indexing multidimensional spaces and is designed for efficient k-NN search~\cite{DES}.


Kd-trees organise $K$-dimensional vectors of numeric data. Each internal node of the tree represents a branching decision in terms of a single attribute's value, called a \textit{split value}~\cite{Talbert}. These internal nodes generate a splitting hyperplane that divides the space into subspaces using this split value, usually the median value along the splitting dimension. We will use the median when constructing the Kd-tree for the data in our experiments. Building a Kd-tree (with the number of dimensions $K$ fixed, and dataset size $N$) has $O(N\log N)$ complexity~\cite{Arya}. For more details on Kd-trees and k-NN search implementations, refer to~\cite{Arya}.

The k-NN search algorithm aims to find a node in the tree closest to a given input vector. Searching a Kd-tree for $k$ nearest neighbours is  $O(k\log N)$, which is the key to fast indexing. It uses tree properties to quickly eliminate large portions of the search space~\cite{DES}.

\section{Methods of Em-K Indexing }

\subsection{Indexing for Deduplication}

Deduplication refers to identifying matching records within a single database and has many applications in database and business contexts. For instance, many businesses maintain databases of customer information that are utilised for advertising purposes, e.g., emailing flyers. Duplicate entries might arise because of errors in data entry or address changes. Deduplication techniques are useful to remove duplicate entries and to improve the quality of the collected information. Duplicate-free customer information databases prevent emailing several copies of flyers to the same customer, which reduces the cost of advertising, but there are many other benefits.

Indexing is a preprocessing step to avoid the need to perform detail-level comparisons between $O(N^2)$ pairs of records. The proposed indexing method utilises the properties of vectors and Kd-trees in a Euclidean space. The method has two main steps,

\begin{enumerate}

    \item Embedding the blocking values: We select the blocking criteria based on the attributes of the records in a dataset. For instance, given a set of records with entity identifying attributes such as first name, last name, date of birth, or postcode, we may choose one or several values of them in our indexing method. We embed the blocking values of a dataset/database in the Euclidean space by applying LSMDS. The embedding depends on the size of the dataset. We propose two techniques:
        \begin{enumerate}
            \item\textit{Complete LSMDS}: For a given dataset of size N, apply LSMDS for the blocking values of all the records.
        
            \item \textit{Landmark LSMDS}: Apply complete LSMDS only to a fraction of the dataset (the landmark records). Then the remaining records are embedded using the out-of-sample embedding against the landmark points. We explained this approach in Section 2.
	
        \end{enumerate}

	\item Nearest neighbour search: Searching for similar points is a two-step procedure.    
	
	    \begin{enumerate}
	        \item 	The first step is to build the Kd-tree in the Euclidean space using all the points that represent the blocking values in this space. 
	
	        \item The second step is to create blocks of similar points by searching the nearest-neighbours of the Kd-tree nodes. 
	        
	    \end{enumerate}

\smallskip	    
Since the Kd-tree construction uses all the available points in the Euclidean space, each record becomes a node in the Kd-tree. Likewise, each node becomes a query against the rest of the nodes in the k-NN search. Each node has a fixed number of nearest neighbours (NNs) allocated for them as we keep the k-NN search fixed for every querying node. Hence, each node in the tree becomes a small block of records that contains its NNs as the members. The block sizes are uniform and depend on the number of NNs allocated for a node.    

\end{enumerate}

Once all the blocks are determined by the indexing method, we return the pairs of similar points in each block as potential matching records. Then, we retrieve the original blocking values that correspond to these points and compare them to classify the pairs as candidate matches or non-matches based on a pre-selected threshold.

A detailed level comparison among the other attribute values will be only required between the pairs that indexing identifies as candidate matches. Thus our indexing method act as a filtering step that reduces the total number of detailed comparisons one has to perform when identifying similar records in a dataset.


\subsubsection{Complexity}

We can quantify the complexity of the proposed indexing method. The method has two components: a relatively slow step where the records (blocking values) are mapped to a Euclidean space, followed by a relatively fast step that creates blocks in the Euclidean space.

Assume that we have $N$ records in the underlying database. Complete LSMDS requires calculating the distance between all pairs of blocking values, hence, $O(N^2)$ operations for the embedding. This complexity dominates the LSMDS calculations. However, for large $N$, we can choose a set of landmarks $L$ and apply LSMDS with a complexity of $O(L^2)$. The embedding of the rest of the points has a linear complexity of $O(ML)$ operations, where $M=N-L$. Hence the overall complexity of landmark LSMDS is $O(L^2 + ML)$.


The second phase builds and searches the Kd-tree to create blocks of records. Building the Kd-tree requires $O(N \log N)$ operations, and searching for $k$ nearest neighbours for $N$ points requires $O(Nk \log N)$ operations, where the size of a block ($B$) is equal to $k$. Hence the overall complexity of the indexing step is $O((1+k)N \log N)$.

Since a complete LSMDS requires $O(N^2)$ operations, we recommend using the landmark LSMDS and as small as possible $k$ to reduce the complexity of the proposed method. 


\subsection{Indexing for Query Matching}

Query matching in this paper refers to querying a stream of records against a reference database to find records that refer to the same entity as the query (see Section 3). Query processing should be quick and accurate for many ER solutions to increase their usability in real-time applications. In some applications, a stream of queries might need to be processed within a given time, collecting as many matches as possible. In such settings, we have to trade accuracy against speed when detecting matching records. This section presents a scalable indexing method for real-time, approximate query matching against a large-scale database. The ideas presented in Section~4.1 serve as the basic building blocks for the proposed method.


Following the \textit{clean-clean ER} scenario, we consider a large-scale reference database $E_r$ and a batch of streaming queries $Q$. Similar to the previous indexing method, we first embed all the blocking values of the reference database in a Euclidean space. We use a set of landmarks and the out-of-sample embedding of LSMDS to reduce the overhead of embedding the database $E_r$. The embedding is a two-step procedure: 1) apply LSMDS over the landmarks 2) then map the rest of the blocking values using out-of-sample embedding of LSMDS based on the distances to these landmarks. We then construct a Kd-tree using all the points in the Euclidean space, where each data point becomes a node in the tree.

For streaming queries, we process a single query record at a time. First, we embed the blocking value of the query record in the pre-mapped Euclidean space. In general, out-of-sample embedding would require calculating all the distances from the new query to the pre-mapped blocking values in the original string space, which is not desirable with large-scale databases. Hence, we only calculate the distances to the landmarks when mapping a new record. The mapping position is determined by applying the out-of-sample embedding of LSMDS to the new query record. The process is similar to mapping the blocking values that are not landmarks in the reference database.

The only inputs we need for the mapping are the distances from a query to the pre-selected landmark blocking values. Then we search for the k-NNs of the new point using the existing Kd-tree structure of the reference database. A new block of points that contains similar points for the query point will be determined. The block size $B$ depends on the $k$ of the k-NN search. We retrieve the original blocking values for the block of similar points and use a pre-selected threshold to filter out the potential matching pairs. The records that contain these as blocking values are the matching records to the querying record. A detailed level comparison between the candidate matching pairs may be required after the initial step of indexing. 


The algorithm needs to process a query within a fixed time, potentially sub-second. Since we are querying a stream of query records against a large-scale database, the processing time of a single query is limited. The overall system performance can be improved by trading accuracy against efficiency and scalability. We will fine-tune our method to trade-off many comparisons against accuracy to offer a greater number of detections within a fixed time. Thus we select a set of optimal parameters that scale our data by considering the trade-off between accuracy and scalability. 


\subsubsection{Complexity}

We can quantify the complexity of different stages of the Em-K indexing for query matching. Similar to the previous method, it contains two components: a relatively slow step where a new query is mapped to a pre-mapped Euclidean space, followed by a relatively fast step that creates a block of similar points which contains potential matches for \mbox{the query}. We assume below that the reference database has already been embedded since this cost is amortised across many queries.

Suppose that we have $N$ data points in our reference database. Typically it would require~$O(N)$ operations to compare the existing records with a query point, which is not feasible for larger~$N$. We avoid this complexity by applying out-of-sample embedding of LSMDS using a set of~$L$ landmarks, which requires only~$O(L)$ operations to embed a new data point. The embedding is efficient if $L$ is chosen such that $L \ll N$.

In the second phase, we search for $k$ nearest neighbours for the query. Therefore, the block size $B$ is equal to $k$. It will cost  $O(k \log N )$ operations to search the tree and $O(k)$ operations to compare a new query point within a block. The total cost of indexing is $O(L+ k \log N)$. However, \mbox{$k \ll L \ll N$} in large-scale applications. The cost of the k-NN search is insignificant due to the efficient indexing structure of the Kd-tree. Thus out-of-sample embedding step dominates the complexity of the proposed method.

The number of landmarks $L$ will affect both the accuracy and scalability of this method. We will discuss the selection of landmarks and results in the next section.

\section{Experimental Validation}

We evaluated the two proposed Em-K indexing methods under various settings (e.g., different dimensions, varying block sizes and datasets, and error rates). Two main questions to study in the experiments are: (1) Are these proposed methods robust over various settings? (2) Does Em-K indexing achieve high accuracy and good scalability?

All algorithms are implemented in R and executed on a desktop with Intel Core 5 Quad 2.3GHz, 16GB RAM, and MacOS Big Sur.

\subsection{Set Up}

\subsubsection{Data Sets}

We examined the performance of our methods over two synthetic datasets. They can be manipulated to have significant variations in their size and characteristics (e.g., error rates).

\begin{itemize}

    \item[-] \textit{\textbf{Dataset-1:}} The first data set contains records with synthetically generated biographic information. Each record has a given name and a surname. They are generated using the tool Geco~\cite{geco}. We introduced duplicate records with errors by slightly modifying the values of randomly selected entries. In record generation, we assumed a record has only one duplicate with a maximum of two typographical errors (substitutions, deletions, insertions, and transpositions) in both attribute values. 

    For the deduplication datasets, there is one duplicate for a particular record within the dataset. Similarly, in query matching, each query has one duplicate within the reference database, and the reference database is duplicate free.

    \item[-] \textit{\textbf{Dataset-2:}} The second dataset is the benchmark dataset presented in~\cite{Saeedi}. It is based on personal records from the North Carolina voter registry and synthetically generated duplicates using Geco~\cite{geco}. Each record has several attribute fields. We cannot control the errors of the duplicate records in this dataset since it has been formulated as a benchmark~dataset. However, after careful analysis of the dataset, we estimated that a duplicate record has a maximum of three edit distance errors for this work. We have only considered the first name and the last name fields in our experiments.

    Similarly, we selected the deduplication data such that there is only one duplicate for a particular record within the dataset. In the query matching, we choose the queries to have only one duplicate within the reference database and the reference database to be duplicate free. 
    
\end{itemize}

\subsubsection{Matching Rates}

We control the number of duplicates in our experiments in order to understand how well the method works in different circumstances. We used two matching rates: Consider the datasets $E$, $E_r$ and $Q$ with sizes $|E|$, $|E_r|$ and~$|Q|$, 

\begin{enumerate}

    \item \textit{Deduplication matching rate ($DMR$)}: the matching rate for deduplication is defined to be the  $DMR(E)= \frac{E_d}{|E|}$ where $E_d$ is the number of duplicate records in $E$.
    
     \item \textit{Query matching rate ($QMR$)}: the matching rate for query matching of $Q$ against a reference database $E_r$ is defined to be the $QMR = \frac{M_{RQ}}{|E_r|}$ where $M_{RQ}$ is the number of records in $E_r$ that has a matching record (duplicate) in $Q$. 
     
\end{enumerate}

We control the number of duplicates within and between datasets by changing the $DMR$ and $QMR$. We represent $DMR$ and $QMR$ as percentages in our experiments. 


\subsubsection{Performance Evaluation Metrics}

We use two measures to quantify the efficiency and the quality of our indexing method proposed for deduplication~\cite{Christen2007}. 


\begin{itemize}

    \item Reduction Ratio (RR): Measures the relative reduction of the comparison space, given by $RR=1- \frac{N_b}{|E| (|E|-1)}$ where $N_b$ is the number of potential matching pairs produced by an indexing algorithm. This quantifies how useful the indexing is at reducing the search space for detailed comparisons.
    
    \item Pair Completeness (PC): This is given by $PC=\frac{N_m}{M}$, where $N_m$ denotes the detected number of real matches by the indexing algorithm and $M$ represents the number of all real matches in $E$. This is a measure of how accurate the indexing is. 
    
\end{itemize}

Both PC and RR are defined in the interval $[0, 1]$, with higher values indicating higher recall and efficiency, respectively. However, PC and RR have a trade-off: more comparisons (higher $N_b$) allow high PC but reduce the RR. Therefore, indexing techniques are successful when they achieve a fair balance between PC and RR.

We used two measures to evaluate the performance of our indexing method of query matching. These are different from the standard measures of indexing defined above as we combine indexing with query matching here. Hence, in this context, we are interested in measuring the efficiency of the method in terms of time and speed of processing a query. 

The Em-K indexing method returns a block of records that contains both true positive (TP) and false positive (FP) matches per query as a final result. Hence the following measures are used.

\begin{itemize}
 
    \item Number of true positives per computational effort: Measures the number of true matching records determined by the indexing method when processing queries within a set period. 

    \item Precision: Measures the accuracy of the query matching in terms of precision. The precision $P$ is denoted by $P=\frac{|TP|}{|TP|+|FP|}$, where $|TP|$ is the number of true positives and $|FP|$ is the number of false positives.

\end{itemize}
All the CPU running times are measured in \textit{seconds} and denoted by RT.

\subsection{Choice of Parameters}

Several factors may impact the performance of the  proposed 
methods; some of them (e.g., dimension ($K$), block size ($B$), landmarks ($L$) ) are control parameters that we rely on to fine-tune the performance, while others (e.g., dataset size or error rate) are parameters determined by data sets.

In this section, We initially discuss the choices of the $K$, $B$, $L$ parameters in our indexing implementations (and their rationale). The robustness of the methods is measured with respect to the varying sizes and matching rates of the data sets in Sections~5.5–5.7.

\begin{figure}[ht!] 
\centering
\includegraphics [height=1.7in, width=3.2in]{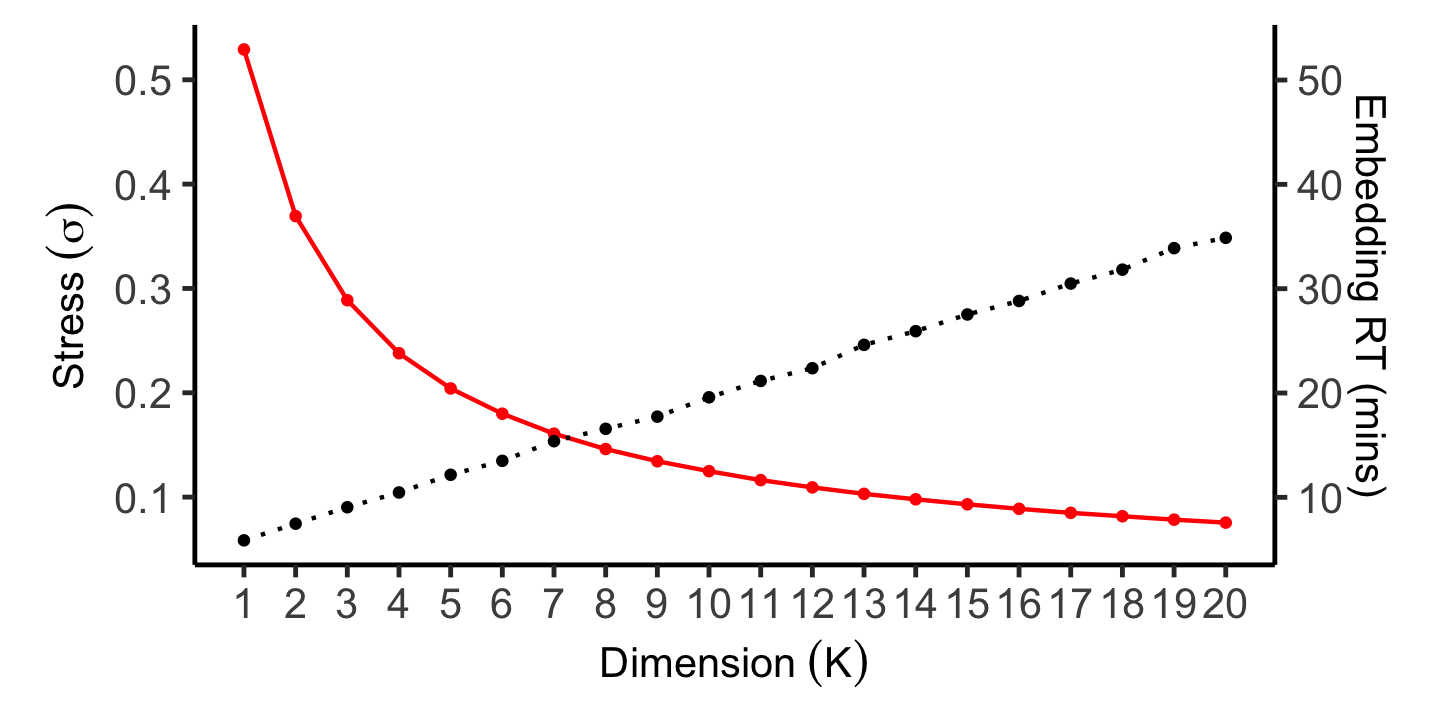}
\caption{The trade-off between the dimension ($K$) vs stress ($\sigma$) and embedding time. The first y-axis represents $\sigma$, while the second y-axis represents the embedding time. The $\sigma$ tends towards a small but non-zero asymptote when $K$ increases. The running time increases linearly when K increases. Higher dimensions allow lower $\sigma$ values for the embedding but increase the embedding time, for marginal benefit.}
  \label{fig:1}
  \Description{ Line plot with two y-axises. The right y-axis shows the stress values ranging from 0.1 to 0.7, and the left y-axis shows embedding running time in minutes. It ranges from 0-50 minutes. The x-axis represents the dimension which varies from 1 to 20.}
\end{figure}

\textbf{$\mathbf K$}: The dimension of the Euclidean space ($K$) plays an important role in the performance of our indexing methods. We applied LSMDS to a sample of 5000 records selected from Dataset-1. In \autoref{fig:1}, the first y-axis shows the stress ($\sigma$ defined in \autoref{eq:1}) decreases as $K$ increases. 

A good value of $K$ should differentiate similar objects from dissimilar ones by approximating the original distances. If $K$ is too small, we will have high-stress values where dissimilar pairs will not fall far enough from each other. It could also place dissimilar pairs closer and similar pairs further apart. Conversely, high $K$ values will have low-stress values. However, in terms of RT, higher dimensions increase the embedding time. In \autoref{fig:1}, the second y-axis represents embedding RT. It takes more than 30 minutes to embed the dataset in 19 or 20 dimensions.



Considering the trade-off between the \textit{stress vs dimension} and the \textit{embedding time vs dimension}, a reasonable value of $\sigma$ is found around 6-8 dimensions. This is consistent with the embedding name strings in the Euclidean space, as discussed in detail in~\cite{herath2020simulating}. We will use $K=7$ here.

{$\mathbf{B}$}: Block size is a dominating factor that directly affects the effectiveness and efficiency of many indexing techniques. Large block sizes increase RT in the indexing step and have a low RR and high PC values. In contrast, small block sizes lead to high RR values with fewer comparisons within each block. However, this may result in low PC values due to missing some matches. Blocks of similar points are determined by k-NN search in Em-K indexing methods. Hence, $B$ is equal to $k$ (number of nearest neighbours). We will consider the choice of $B$ in detail in Section~5.2.1.

\textbf{$\mathbf L$}: The number of landmarks ($L$) is another important factor in our indexing methods. Landmarks are utilised for two different purposes in this work. First, we use landmarks for embedding large-scale data into a Euclidean space when the standard LSMDS method becomes inefficient. Second, we use landmarks to embed the out-of-sample queries in a pre-mapped Euclidean space. We discuss the role of landmarks in deduplication and their impact on the proposed indexing method in Section~5.2.2.

In the following experiments we investigated how $B$ and $L$ impact our indexing algorithms.

\subsubsection{Varying Block Sizes ($B$)}

To test the choice of $B$, we used sample datasets containing 5000 records from the two data sources. We set the $DMR=10$\% for the data selected from Dataset-1, which means there are 500 duplicates within the selected 5000 records in the sample dataset. Similarly, we used $DMR=7.5$\% for the second data sample selected from Dataset-2. Hence within the 5000 records, 375 of them are duplicates.

\begin{figure}[ht!] 
\centering
\includegraphics [height=1.8in, width=3.3in]{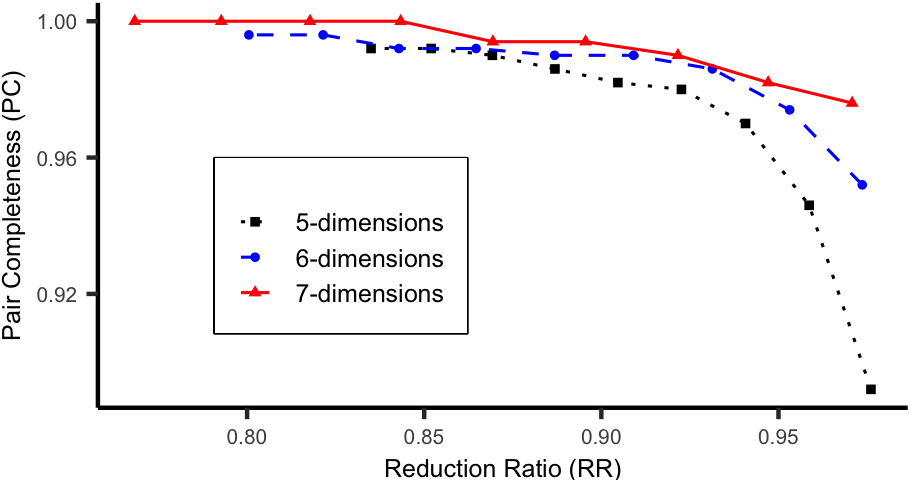}
\caption{The trade-off between the reduction ratio (RR) and pair completeness (PC) in different dimensions. Ideally, RR=PC=1, hence we prefer methods whose results lie as close to the top-right corner of the graph. The block sizes are 100, 90, 80, 70, 60, 50, 40, 30, 20. Large blocks achieve higher PC but lower RR. The three curves illustrate that the higher the dimension, the better the results, up to the point of diminishing returns.}
\label{fig:2}
\Description{ A line plot with three lines that represent results of different dimensions. The y-axis contains pair completeness values ranging from 0.9 to 1. The x-axis represents the reduction ratio values with a range of 0.7 to 1.}
\end{figure}

We performed deduplication indexing on the two datasets. \autoref{fig:2} illustrates the trade-off between PC and RR of our indexing method for different block sizes using the Dataset-1. In each instance, we changed $B$ by varying $k$ in the k-NN search. PC increases with the increase of $B$. In contrast, RR decreases due to the increment in the number of comparisons within each block of records.

The results also indicate that dimension around 7 are good at shifting the PC-RR curve to the top-right corner delivering a good ratio between RR and PC. However, higher dimensions also mean higher computational costs (e.g., high RT) for the embedding. Based on \autoref{fig:1} and \autoref{fig:2}, therefore, we conclude that using $K=7$ and $B=\{50, 60\}$ gives the best compromise PC-RR ratio and RT overall for the given data set. In subsequent experiments, we used $K=7$.

\begin{figure}[ht!] 
\centering
\includegraphics [height=1.4in, width=2.6in]{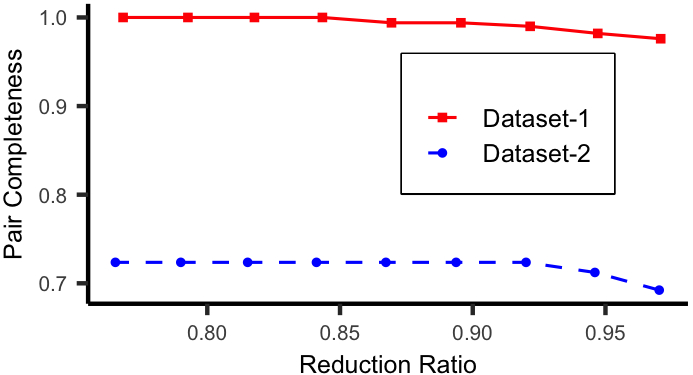}
\caption{The trade-off between the reduction ratio (RR) and pair completeness (PC) for two different datasets for the proposed indexing. The block sizes are 100, 90, 80, 70, 60, 50, 40, 30, 20. PC is very low for the second dataset while RR values are closer.}
\label{fig:3}

\Description{ A line plot with two lines that represent results of different datasets. The y-axis contains pair completeness values ranging from 0.7 to 1. The x-axis represents the reduction ratio values with a range of 0.7 to 1.}

\end{figure}

In most comparisons, we observed similar results for both datasets and discussed only the results of Dataset-1. However, we present one comparison that has comparably different results here. \autoref{fig:3} compares the two datasets in a fixed dimension ($K=7$), varying $B$ in each instance. Both data sets achieve similar RR values with different PC values. PC values are comparatively low for Dataset-2 and are around 70\% for most occurrences.

Dataset 1 shows better results compared to Dataset-2 in \autoref{fig:3}. This behaviour is expected since the two datasets have different characteristics, e.g., different matching rates and a different number of errors in each field. Furthermore, we used different pre-selected thresholds ($\theta_m$) when validating candidate matching pairs in the two datasets. These thresholds are selected based on the errors in the two datasets. In Dataset-1, duplicates have a maximum of two typographical errors and therefore $\theta_m=2$. For Dataset-2, we assumed that each duplicate record has a maximum of three typographical errors, and we set $\theta_m=3$ therein.


\subsubsection{The Effect of Landmarks}

The following experiment investigated the effect of the two different embedding techniques on the proposed indexing method. Performance is measured using the PC and RR curves. First, we applied complete LSMDS similar to \autoref{fig:2}, keeping the $K=7$ fixed. Second, we applied LSMDS only to a set of landmarks in the same dimension. The remaining points are embedded using the out-of-sample embedding of LSMDS and the distances to landmarks. Blocksize $B$ is varied as before.

\begin{figure}[ht!] 
\centering
\includegraphics [height=2.1in, width=4.1in]{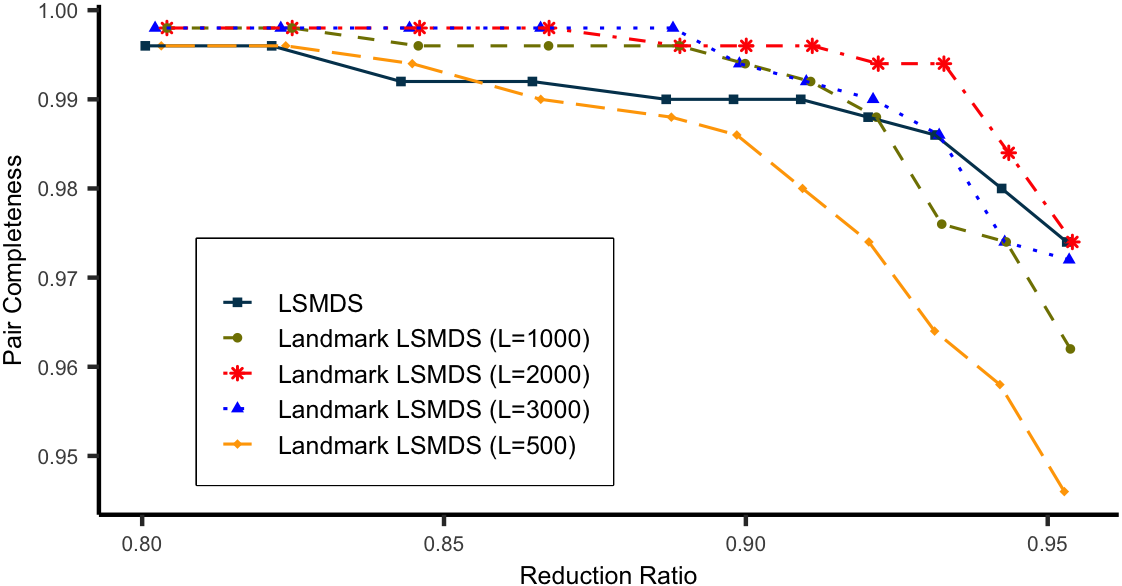}
\caption{The trade-off between the reduction ratio (RR) and pair completeness (PC) for the proposed indexing based on different embeddings.The results are based on a complete LSMDS and landmark based LSMDS for different number of landmarks. The block sizes are 100, 90, 80, 70, 60, 55, 50, 45, 40, 35, 30.}
\label{fig:4}

\Description{ A line plot with five lines that represent results of different LSMDS embeddings. The y-axis contains pair completeness values ranging from 0.95 to 1. The x-axis represents the reduction ratio values with a range of 0.8 to 1.}

\end{figure}


\autoref{fig:4} compares the trade-off between PC and RR for complete LSMDS and landmark LSMDS (for different $L$). Each instance represents a different block size. PC and RR change similarly to the previous experiment for different block sizes. However, \autoref{fig:4} suggests that we can get similar results by choosing an approximate embedding that uses landmarks instead of complete LSMDS. The use of landmarks decreases the distance calculations and we can avoid the inherent complexity and inefficiency of LSMDS when processing large-scale data.



Using our indexing solution, we can solve deduplication applications in ER. The method requires embedding records into a Euclidean space in order to apply the indexing technique. Since complete LSMDS is not suitable for large-scale data, we recommend using the landmark LSMDS. We applied the farthest first sampling~\cite{KAMOUSI20161} for reproducible results in landmarks selection; however, random selection works well in practice. 

The optimal parameter setting for our data is $K=7$, $B= 50$ and $L=1500$. We used the proposed indexing method of deduplication as a motivating example for the next set of experiments.

\subsection{Indexing for Query Matching}

In our experiments, we used a reference database with $E_r=5000$ records. The streaming query dataset $Q$ is flexible in size because we only consider streaming queries within a period. Each query in $Q$ has a duplicate record in the reference database, i.e., $QMR=1$. We made this assumption to keep the experiment more efficient instead of mimicking a real-world ER problem. In a real-world scenario, each query may not have a matching record within the reference database, or the same query may appear in the stream to search against the reference database. However, the time required for searching is the same when no match is present. 



Similar to the previous method, several factors affect the performance of the proposed indexing method for querying, e.g., $K, L, B$. Since the embedding of the reference database is the same as before, we keep $K=7$. The block size is $B$ is equal to $k$, the number of nearest neighbours in k-NN search.

We used landmarks to support the out-of-sample embedding. The number of landmarks, $L$, directly impacts the running time (RT) of the proposed method since each query needs to be embedded in the Euclidean space. The other costs that contribute to RT are the distance calculations and k-NN search.

First, we embedded the reference database generated from Dataset-1 applying landmark LSMDS. We chose the landmarks based on the farthest-first sampling. Then, we built a Kd-tree using all the reference data points in the Euclidean space. Once the Kd-tree is built, we passed queries to search the tree for k-NNs. Hence, each query in $Q$ needs to be mapped in the Euclidean space. We used the same set of landmarks among the reference data points to map the query points. Distance calculations are required among the new query point and the landmarks when applying the out-of-sample embedding of LSMDS. A block of similar points is determined for a new query by searching k-NNs in the existing Kd-tree.

\begin{figure}[ht!] 
\centering
\includegraphics [height=2.1in, width=4.1in]{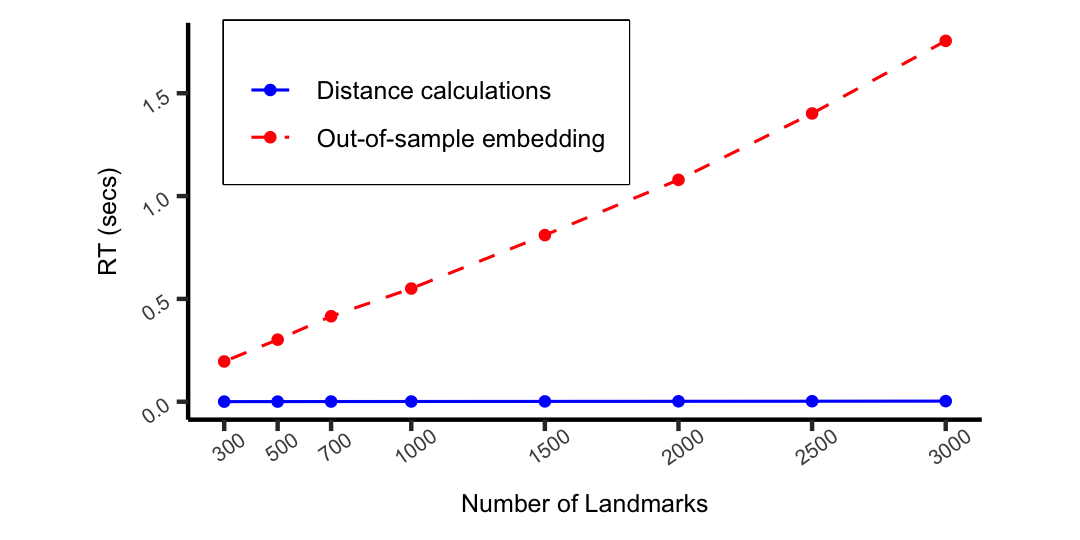}
\caption{The calculation times for distance calculations and out-of-sample embedding for query matching. Both depend on the number of landmarks, but distance calculations are much faster.}
\label{fig:5}

\Description{ A line plot with two lines that represent running times. The y-axis contains running times ranging from 0 to 2 measured in seconds. The x-axis represents the number of landmarks with a range of 300 to 3000.}

\end{figure}

Our method process a single query at a time. In the following experiment, we processed 500 queries against the reference database. For each query, we measured the embedding RT and the distance calculation RT separately. Then we calculated the mean values for both categories, processing all the 500 queries. \autoref{fig:5} shows the comparison of the mean RT of distance calculations and out-of-sample embedding for varying numbers of landmarks. Increasing $L$ linearly increases the embedding RT of a single query. The distance calculation RT also increases linearly with $L$ but is negligible compared to out-of-sample embedding RT.

We also measured the cost of the k-NN search when creating a block of records for a new query. This search can be done efficiently in the Euclidean space using the Kd-tree and priority queues. It takes less than a millisecond, which is insignificant compared to the total RT of the embedding process. Moreover, increasing $k$ has a smaller impact due to its efficient implementation with priority queues~\cite{Arya}.

A scalable query matching method should be able to process as many queries as possible within a period. In our indexing method, increasing $L$ limits the number of queries processed within a set period. On the other hand, small $L$ tends to decrease the accuracy of the embedding. As a result, a new query may be not mapped closer to its duplicate, reducing the probability of grouping them as similar points. Hence an optimal $L$ is required to maintain the scalability without degrading the quality of the results.

We measured the scalability and the efficiency of our method using two quantitative measures with respect to time: the number of true positive ($|TP|$) matches detected per computational effort and the precision (P) per computational effort. Hence we processed a stream of queries $Q$ against a reference database $E_r$ within a given period, varying the control parameters such as $L$ and $B$ in different instances. Then we calculated $|TP|$ and $|FP|$ found by our method in each instance. The accuracy of the results is measured in terms of precision. It measures the rate of TP against all the positive results (sum of $|TP|$ and $|FP|$) returned by the method within the given period. Subsequently, we determined an optimal set of parameter values for our data that returns the highest TP matches and precision within a fixed period.

In the following experiments, we used a reference database of $E_r=5000$ records and the stream of query records $Q=500$. We applied landmark LSMDS to embed the records in $E_r$ to the Euclidean space in each experiment. Then the queries are processed using the same set of landmarks. Hence, every instance of a different $L$ has a disparate embedding of the reference database, then used for query embedding and searching. 


\begin{figure}[ht!] 
\centering
\includegraphics [height=2.1in, width=4.1in]{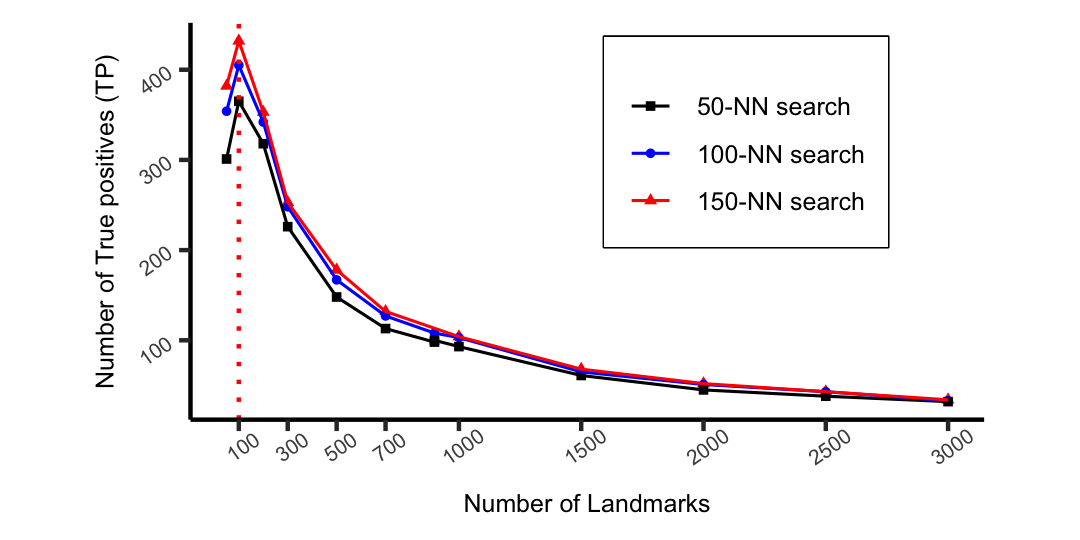}
\caption{The trade-off between the number of true positives and the number of landmarks for three different block sizes. Here the block size is equal to the number of k-NNs. Many landmarks allow fewer queries to be processed, while many k-NNs allow more true match detections.}
\label{fig:6}

\Description{ A line plot with three lines that represent results of a different nearest neighbour search. The y-axis contains the number of true positive values ranging from 0 to 500. The x-axis represents the number of landmarks with a range of 300 to 3000.}

\end{figure}

\autoref{fig:6} shows the comparison of the $|TP|$ against varying the number of landmarks and k-NNs. In each instance, we processed queries within a fixed period (60 seconds). Increasing $L$ decreases~$|TP|$ because more landmarks allow fewer queries to be processed. This phenomenon is expected since large $L$ increase the running time of processing a single query. With more landmarks, there is a higher probability of finding matches for those queries due to the increasing accuracy of the embedding. In contrast, few landmarks allow more queries to be processed within a period since the running time (RT) for embedding a single query is small. 


\autoref{fig:6} suggests that we only need 100 landmarks to detect the highest number of TP matches for the given data. The method has processed all the 500 queries within a minute, detecting 432 TP matches. The average time for processing a single query is 0.07 seconds. 



Based on \autoref{fig:6}, we conclude that setting $k=150$  and $L={100}$ gives the best trade-off between the quality of the results and the RT. This result is consistent with existing real-time query matching techniques that process a query within a sub-second time~\cite{Liang}.

\begin{figure}
  \begin{subfigure}{10cm}
    \centering\includegraphics[width=7.8cm]{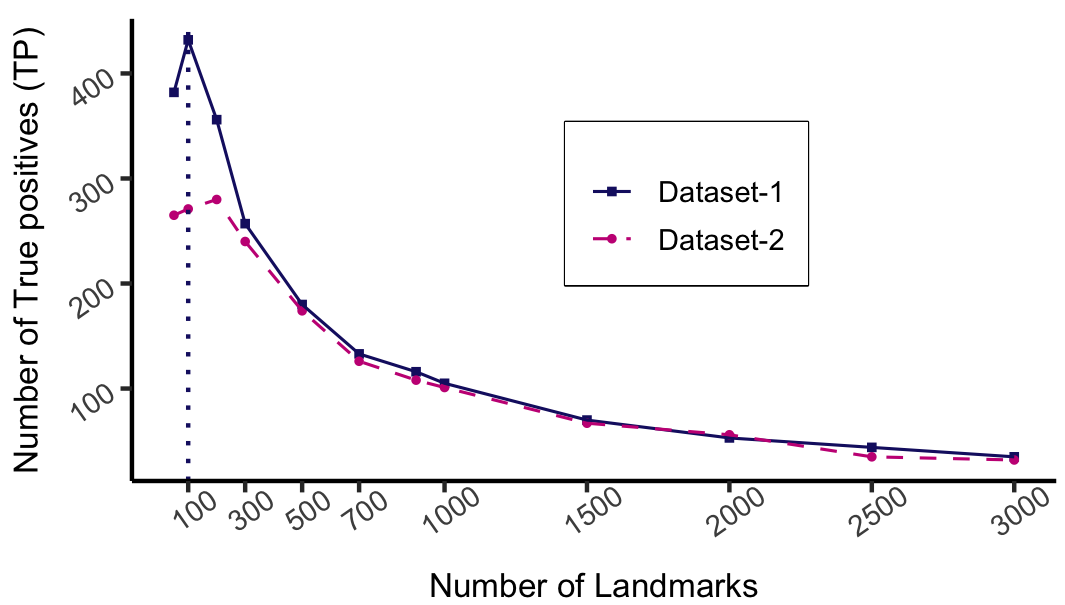}
    \caption{The trade-off between the number of TP and the number of landmarks for two different datasets.}
  \end{subfigure}
 
  \begin{subfigure}{10cm}
    \centering\includegraphics[width=7.8cm]{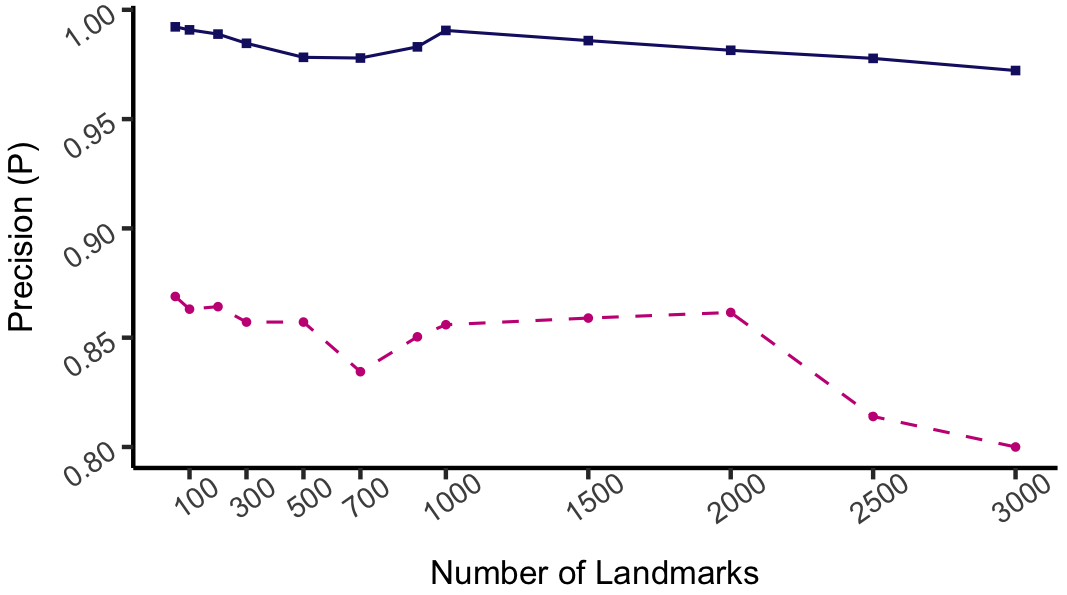}
    \caption{The trade-off between the precision and the number landmarks for two different datasets.}
  \end{subfigure}
  
    \caption{The two figures (a) and (b) compares the Dataset-1 and Dataset-2 in terms of number of TP and Precision varying the $L$  The k-NNs are fixed to $k=150$ and $T=60$ seconds in each instance. The curves illustrate similar trends for similar parameter settings in (a). However, in (c), Dataset-2 exhibits low precision compared to Dataset-1.}

  \label{fig:7}
  
  \Description{Two line plots a and b displaying the precision and the number of true positives against different numbers of landmarks. The y-axis of plot a has a value range from 100 to 500. The y-axis of the second plot range from 0.8 to 1. Both plots have the same x-axis with a range of 100-3000 landmarks. Each plot has two curves in different colours.}

\end{figure}


To validate the robustness of the method, we compared the results of applying the EM-K indexing method to two reference databases and stream of queries derived from Dataset-1 and \mbox{Dataset-2}, respectively in \mbox{\autoref{fig:7}}. Both reference databases contained 5000 records. The two streams of queries are processed against the two reference databases separately with similar parameter settings. The query matching rate ($QMR$) is equal to 1 for both.  Keeping the control parameters fixed at $K=7$, $B=150$ and $T=60$ seconds, we vary $L$ to compare different results.

\mbox{\autoref{fig:7}-(a)} compares the trade-off between the $|TP|$ per computational effort, and the number of landmarks. Increasing $L$ decreases the $|TP|$ for query matching in both datasets similar to \autoref{fig:6}. Few landmarks allow processing all the queries within 60 seconds, whereas many landmarks allow processing fewer queries. While both reference databases are similar in size, the total queries processed are different in size. A total of 500 queries are processed against Dataset-1, and 375 queries are processed against Dataset-2 in the experiment. Hence, we observe fewer TP matches for Dataset-2 than Dataset-1 in their highest performance. However, the curves illustrate similar trends for both datasets.


Based on the results produced by the previous experiment, \mbox{\autoref{fig:7}-(b)} compares the two datasets in terms of the precision (P) against varying $L$  In each instance, the queries are processed within T=60 seconds. Hence the P values are measured per computational effort. Dataset-2 has lower P values compared to Dataset-1. This behaviour is expected since we assume that the duplicate records in Dataset-2 contain a maximum of three edit distance errors. The pre-selected thresholds~($\theta_m$) are different for the two datasets. However, we do not have the exact details of error rates for this benchmark dataset. Hence, the method may allow more false positives in the final block of records retrieved for a given query, reducing the P.

However, the overall results suggest that the proposed method is robust over different datasets. The optimal parameter set that provides the best compromise for our data is  $K=7$, $B$ (or $k$)$= 150$ and $L={100,300}$. We need a few landmarks to achieve the most number of TP matches for a stream of queries within a period. Few landmarks make the embedding efficient. As a result, a single query can be processed within less than a second against a reference database to find matches. Hence our method scales well for approximate query matching against a large-scale reference database for ER.


\section{Discussion}

The Em-K indexing methods embed a set of strings in a metric-space, particularly a lower-dimensional Euclidean space. There is a trade-off between the dimension and the accuracy of the embedding. Higher dimensions allow the embedding to be more accurate. However, it does not scale well for large datasets. In our experiments, we only selected two blocking variables for embedding and indexing. We can also use other blocking variables, such as addresses and gender. As a result, the dimension of the Euclidean space needs to extend accordingly to facilitate those.

The most costly part of the method is the amount of time to embed queries, which increases linearly with the number of landmarks. However, this approach is easily parallelizable since each query is processed separately. We can tune the parameters for a fast, less accurate query matching or a slow, more accurate method depending on the application. 


Our method is designed to solve approximate query matching rather than exact query matching. It means we expect the queries to contain errors in their attribute values. We could easily perform an exact search based on lexicographical order considering the query and the reference dataset using a data structure such as a binary tree to find exact matches as a pre-filter for our method.


Our indexing method for query matching uses distance computations to embed the reference database in the Euclidean space. This embedding is a slow process that requires a minimum of $O(L^2)$ operations where $L$ is the number of landmarks. We consider this as the training phase, which only needs to be performed once. However, with the Kd-tree built, we can perform similarity queries with significantly fewer distance computations than a sequential scan of the entire dataset.

In some applications, large-scale reference databases grow with time. Hence the new entities need to be added incrementally. For example, some applications would require the addition of queries that have no matching records in the reference database already. To facilitate those, we need to extend the Kd-tree accordingly without repeating the embedding process that creates the initial tree. However, growing a Kd-tree can be a heuristic procedure where the tree could become unbalanced. Therefore, we can explore alternative tree structures such as R-tree that are robust against dynamic data.


\section{Related Work}

Indexing techniques are recognized as a crucial component for improving the efficiency and the scalability	of the ER solutions. There exist a variety of indexing techniques which are also known as blocking or filtering. \mbox{Papadakis \textit{et al.}~\cite{sur2}} and Christen \textit{et al.}~\cite{sur1} presented surveys that include well-known indexing methods such as standard blocking, suffix array, q-gram blocking, sorted-neighbourhood, canopy clustering, and string-map based methods. We only survey a few closely related works since many of these existing methods are orthogonal to the focus of this paper.

Canopy clustering uses cheap comparison metrics to group similar records into overlapping canopies and creates blocks from records that share a common-canopy~\cite{MacCallum}. The method depends on global threshold values, and that reduces its flexibility. It also uses similarity measures such as TF-IDF and Cosine distance that can be computationally expensive~\cite{sur1}.

Mapping-based indexing methods map records to objects in a Euclidean space, preserving the original distances between them. Jin \textit{et al.}~\cite{Li-chen} proposed the Stringmap algorithm that maps records into a similarity preserving Euclidean space (with dimension between 15-20). Similar pairs are determined by building an R-tree. Stringmap has linear complexity, but it requires tuning several parameters. Moreover, the performance of such approaches tends to decrease with more than 20 dimensions. In contrast, our method operates in much smaller dimensions.

The Double Embedding scheme \cite{DES} uses two-dimensions, $K$ and $K'$ for embedding records such that $K'<K$. Similarity joins are performed in the metric-space using a Kd-tree and nearest-neighbour search to find candidate matches. The method is faster than the Stringmap algorithm. However, it attempts to keep the embedding contractive by increasing the distance computations and is not suitable for large-scale data.

Another metric-space indexing technique utilised an M-tree to produce complete and efficient ER results. The cost of the method and the quality of the results have remained similar to existing indexing techniques~\cite{Akgun}. However, it does not scale for large-scale data since it has a single step that combines the indexing, comparison, and classification steps.

In summary, the existing mapping-based indexing techniques have two components. The first is to map the records into a metric-space. The second is to perform similarity joins in the metric-space using a tree-like indexing structure.

These methods were developed offline by applying a mapping technique to map all the records into a multidimensional metric-space. The spatial mapping of records acts as a filtering step before the actual record matching. Hence, the focus is on matching similar records without grouping them into blocks. However, none of them handles new records by reusing the properties of an existing multidimensional metric-space. Therefore,  query matching is not achievable. In this paper, similar ideas are significantly extended to accommodate the benefits of the Euclidean space for efficient query matching.

\section{Conclusions}

Indexing techniques reduce the pairwise comparisons in ER solutions. 
Many existing mapping-based indexing techniques work in offline mode with fixed-size databases. Hence, these techniques are not suitable for applications that require real-time query matching, especially if it involves big, fast, or streaming data. Our method investigated the query matching problem in ER by using spatial mapping of records into a Euclidean space. We aimed to develop an indexing approach for a fast query process within a short time, returning as many as potential matches. The proposed method proved fast running time as well as scalability along with the data size. The use of vectors in the Euclidean space that represent records allowed fast query matching with comparable accuracy. Many directions are ahead for future work. First, we plan to extend the Em-K indexing method to be parallel. Second, the current Em-K indexing can be extended to other forms of ER problems, such as querying against a dynamic database or iterative ER~\cite{HARRA}.



\bibliographystyle{ACM-Reference-Format}
\bibliography{sample-base}


\begin{thebibliography}{32}


\ifx \showCODEN    \undefined \def \showCODEN     #1{\unskip}     \fi
\ifx \showDOI      \undefined \def \showDOI       #1{#1}\fi
\ifx \showISBNx    \undefined \def \showISBNx     #1{\unskip}     \fi
\ifx \showISBNxiii \undefined \def \showISBNxiii  #1{\unskip}     \fi
\ifx \showISSN     \undefined \def \showISSN      #1{\unskip}     \fi
\ifx \showLCCN     \undefined \def \showLCCN      #1{\unskip}     \fi
\ifx \shownote     \undefined \def \shownote      #1{#1}          \fi
\ifx \showarticletitle \undefined \def \showarticletitle #1{#1}   \fi
\ifx \showURL      \undefined \def \showURL       {\relax}        \fi
\providecommand\bibfield[2]{#2}
\providecommand\bibinfo[2]{#2}
\providecommand\natexlab[1]{#1}
\providecommand\showeprint[2][]{arXiv:#2}

\bibitem[\protect\citeauthoryear{Adly}{Adly}{2009}]%
        {DES}
\bibfield{author}{\bibinfo{person}{Noha Adly}.}
  \bibinfo{year}{2009}\natexlab{}.
\newblock \showarticletitle{Efficient record linkage using a double embedding
  scheme.}. In \bibinfo{booktitle}{\emph{DMIN}}, Vol.~\bibinfo{volume}{48}.
  \bibinfo{pages}{274--281}.
\newblock


\bibitem[\protect\citeauthoryear{Akgün, Dearle, Kirby, and Christen}{Akgün
  et~al\mbox{.}}{2018}]%
        {Akgun}
\bibfield{author}{\bibinfo{person}{Özgür Akgün}, \bibinfo{person}{Alan
  Dearle}, \bibinfo{person}{Graham Kirby}, {and} \bibinfo{person}{Peter
  Christen}.} \bibinfo{year}{2018}\natexlab{}.
\newblock \bibinfo{booktitle}{\emph{Using metric space indexing for complete
  and efficient record linkage}}.
\newblock \bibinfo{pages}{89--101}.
\newblock
\showISBNx{978-3-319-93039-8}
\urldef\tempurl%
\url{https://doi.org/10.1007/978-3-319-93040-4_8}
\showDOI{\tempurl}


\bibitem[\protect\citeauthoryear{Arya, Mount, Netanyahu, Silverman, and
  Wu}{Arya et~al\mbox{.}}{1998}]%
        {Arya}
\bibfield{author}{\bibinfo{person}{Sunil Arya}, \bibinfo{person}{David~M.
  Mount}, \bibinfo{person}{Nathan~S. Netanyahu}, \bibinfo{person}{Ruth
  Silverman}, {and} \bibinfo{person}{Angela~Y. Wu}.}
  \bibinfo{year}{1998}\natexlab{}.
\newblock \showarticletitle{An optimal algorithm for approximate nearest
  neighbor searching fixed dimensions}.
\newblock \bibinfo{journal}{\emph{J. ACM}} \bibinfo{volume}{45},
  \bibinfo{number}{6} (\bibinfo{date}{Nov.} \bibinfo{year}{1998}),
  \bibinfo{pages}{891–923}.
\newblock
\showISSN{0004-5411}
\urldef\tempurl%
\url{https://doi.org/10.1145/293347.293348}
\showDOI{\tempurl}


\bibitem[\protect\citeauthoryear{Chatfield and Collins}{Chatfield and
  Collins}{1981}]%
        {JChatfied1980}
\bibfield{author}{\bibinfo{person}{Chris Chatfield} {and}
  \bibinfo{person}{Alexander Collins}.} \bibinfo{year}{1981}\natexlab{}.
\newblock \bibinfo{booktitle}{\emph{Introduction to multivariate analysis}}.
\newblock \bibinfo{publisher}{CRC Press}.
\newblock


\bibitem[\protect\citeauthoryear{Ch\'{a}vez, Navarro, Baeza-Yates, and
  Marroqu\'{\i}n}{Ch\'{a}vez et~al\mbox{.}}{2001}]%
        {Edgar}
\bibfield{author}{\bibinfo{person}{Edgar Ch\'{a}vez}, \bibinfo{person}{Gonzalo
  Navarro}, \bibinfo{person}{Ricardo Baeza-Yates}, {and}
  \bibinfo{person}{Jos\'{e}~Luis Marroqu\'{\i}n}.}
  \bibinfo{year}{2001}\natexlab{}.
\newblock \showarticletitle{Searching in metric spaces}.
\newblock \bibinfo{journal}{\emph{ACM Comput. Surv.}} \bibinfo{volume}{33},
  \bibinfo{number}{3} (\bibinfo{date}{Sept.} \bibinfo{year}{2001}),
  \bibinfo{pages}{273–321}.
\newblock
\showISSN{0360-0300}
\urldef\tempurl%
\url{https://doi.org/10.1145/502807.502808}
\showDOI{\tempurl}


\bibitem[\protect\citeauthoryear{Chen, Chung, Xu, Wang, Qin, and Chau}{Chen
  et~al\mbox{.}}{2004}]%
        {Chen_Chung}
\bibfield{author}{\bibinfo{person}{H. Chen}, \bibinfo{person}{W. Chung},
  \bibinfo{person}{J.J. Xu}, \bibinfo{person}{G. Wang}, \bibinfo{person}{Y.
  Qin}, {and} \bibinfo{person}{M. Chau}.} \bibinfo{year}{2004}\natexlab{}.
\newblock \showarticletitle{Crime data mining: a general framework and some
  examples}.
\newblock \bibinfo{journal}{\emph{Computer}} \bibinfo{volume}{37},
  \bibinfo{number}{4} (\bibinfo{year}{2004}), \bibinfo{pages}{50--56}.
\newblock
\urldef\tempurl%
\url{https://doi.org/10.1109/MC.2004.1297301}
\showDOI{\tempurl}


\bibitem[\protect\citeauthoryear{Christen}{Christen}{2012a}]%
        {Christen2012}
\bibfield{author}{\bibinfo{person}{Peter Christen}.}
  \bibinfo{year}{2012}\natexlab{a}.
\newblock \bibinfo{booktitle}{\emph{Further topics and research directions}}.
\newblock \bibinfo{publisher}{Springer Berlin Heidelberg},
  \bibinfo{address}{Berlin, Heidelberg}, \bibinfo{pages}{209--228}.
\newblock
\showISBNx{978-3-642-31164-2}
\urldef\tempurl%
\url{https://doi.org/10.1007/978-3-642-31164-2_9}
\showDOI{\tempurl}


\bibitem[\protect\citeauthoryear{Christen}{Christen}{2012b}]%
        {sur1}
\bibfield{author}{\bibinfo{person}{Peter Christen}.}
  \bibinfo{year}{2012}\natexlab{b}.
\newblock \showarticletitle{A Survey of indexing techniques for scalable record
  linkage and deduplication}.
\newblock \bibinfo{journal}{\emph{IEEE Transactions on Knowledge and Data
  Engineering}} \bibinfo{volume}{24}, \bibinfo{number}{9}
  (\bibinfo{year}{2012}), \bibinfo{pages}{1537--1555}.
\newblock


\bibitem[\protect\citeauthoryear{Christen and Goiser}{Christen and
  Goiser}{2007}]%
        {Christen2007}
\bibfield{author}{\bibinfo{person}{Peter Christen} {and} \bibinfo{person}{Karl
  Goiser}.} \bibinfo{year}{2007}\natexlab{}.
\newblock \bibinfo{booktitle}{\emph{Quality and complexity measures for data
  linkage and deduplication}}.
\newblock \bibinfo{publisher}{Springer Berlin Heidelberg},
  \bibinfo{address}{Berlin, Heidelberg}, \bibinfo{pages}{127--151}.
\newblock
\showISBNx{978-3-540-44918-8}
\urldef\tempurl%
\url{https://doi.org/10.1007/978-3-540-44918-8_6}
\showDOI{\tempurl}


\bibitem[\protect\citeauthoryear{Christen and Vatsalan}{Christen and
  Vatsalan}{2013}]%
        {geco}
\bibfield{author}{\bibinfo{person}{Peter Christen} {and}
  \bibinfo{person}{Dinusha Vatsalan}.} \bibinfo{year}{2013}\natexlab{}.
\newblock \showarticletitle{Flexible and extensible generation and corruption
  of personal data}. In \bibinfo{booktitle}{\emph{Proceedings of the 22nd ACM
  International Conference on Information and Knowledge Management}} (San
  Francisco, California, USA) \emph{(\bibinfo{series}{CIKM '13})}.
  \bibinfo{publisher}{Association for Computing Machinery},
  \bibinfo{address}{New York, NY, USA}, \bibinfo{pages}{1165–1168}.
\newblock
\showISBNx{9781450322638}
\urldef\tempurl%
\url{https://doi.org/10.1145/2505515.2507815}
\showDOI{\tempurl}


\bibitem[\protect\citeauthoryear{Draisbach and Naumann}{Draisbach and
  Naumann}{2011}]%
        {Naumann}
\bibfield{author}{\bibinfo{person}{Uwe Draisbach} {and} \bibinfo{person}{Felix
  Naumann}.} \bibinfo{year}{2011}\natexlab{}.
\newblock \showarticletitle{A generalization of blocking and windowing
  algorithms for duplicate detection}.
\newblock \bibinfo{journal}{\emph{Proceedings - 2011 International Conference
  on Data and Knowledge Engineering, ICDKE 2011}}, \bibinfo{pages}{18 -- 24}.
\newblock
\urldef\tempurl%
\url{https://doi.org/10.1109/ICDKE.2011.6053920}
\showDOI{\tempurl}


\bibitem[\protect\citeauthoryear{Faloutsos and Lin}{Faloutsos and Lin}{1995}]%
        {FastMap}
\bibfield{author}{\bibinfo{person}{Christos Faloutsos} {and}
  \bibinfo{person}{King-Ip Lin}.} \bibinfo{year}{1995}\natexlab{}.
\newblock \showarticletitle{FastMap: a fast algorithm for indexing, data-mining
  and visualization of traditional and multimedia datasets}. In
  \bibinfo{booktitle}{\emph{Proceedings of the 1995 ACM SIGMOD International
  Conference on Management of Data}} (San Jose, California, USA)
  \emph{(\bibinfo{series}{SIGMOD '95})}. \bibinfo{publisher}{Association for
  Computing Machinery}, \bibinfo{address}{New York, NY, USA},
  \bibinfo{pages}{163–174}.
\newblock
\showISBNx{0897917316}
\urldef\tempurl%
\url{https://doi.org/10.1145/223784.223812}
\showDOI{\tempurl}


\bibitem[\protect\citeauthoryear{Groenen and Velden}{Groenen and
  Velden}{2016}]%
        {MDS}
\bibfield{author}{\bibinfo{person}{Patrick Groenen} {and}
  \bibinfo{person}{Michel Velden}.} \bibinfo{year}{2016}\natexlab{}.
\newblock \showarticletitle{Multidimensional scaling by majorization: a
  review}.
\newblock \bibinfo{journal}{\emph{Journal of Statistical Software}}
  \bibinfo{volume}{73} (\bibinfo{date}{09} \bibinfo{year}{2016}).
\newblock
\urldef\tempurl%
\url{https://doi.org/10.18637/jss.v073.i08}
\showDOI{\tempurl}


\bibitem[\protect\citeauthoryear{Herath, Roughan, and Glonek}{Herath
  et~al\mbox{.}}{2021}]%
        {herath2020simulating}
\bibfield{author}{\bibinfo{person}{Samudra Herath}, \bibinfo{person}{Matthew
  Roughan}, {and} \bibinfo{person}{Gary Glonek}.}
  \bibinfo{year}{2021}\natexlab{}.
\newblock \showarticletitle{Generating name-Like vectors for testing
  large-scale entity resolution}.
\newblock \bibinfo{journal}{\emph{IEEE Access}}  \bibinfo{volume}{9}
  (\bibinfo{year}{2021}), \bibinfo{pages}{145288--145300}.
\newblock
\urldef\tempurl%
\url{https://doi.org/10.1109/ACCESS.2021.3122451}
\showDOI{\tempurl}


\bibitem[\protect\citeauthoryear{Hjaltason and Samet}{Hjaltason and
  Samet}{2003}]%
        {Hjaltason}
\bibfield{author}{\bibinfo{person}{Gisli Hjaltason} {and}
  \bibinfo{person}{Hanan Samet}.} \bibinfo{year}{2003}\natexlab{}.
\newblock \showarticletitle{Index-driven similarity search in metric spaces}.
\newblock \bibinfo{journal}{\emph{ACM Transactions on Database Systems (TODS)}}
   \bibinfo{volume}{28} (\bibinfo{date}{12} \bibinfo{year}{2003}),
  \bibinfo{pages}{517--580}.
\newblock
\urldef\tempurl%
\url{https://doi.org/10.1145/958942.958948}
\showDOI{\tempurl}


\bibitem[\protect\citeauthoryear{Jaro}{Jaro}{1989}]%
        {Matthew}
\bibfield{author}{\bibinfo{person}{Matthew~A. Jaro}.}
  \bibinfo{year}{1989}\natexlab{}.
\newblock \showarticletitle{Advances in record-linkage methodology as applied
  to matching the 1985 census of Tampa, Florida}.
\newblock \bibinfo{journal}{\emph{J. Amer. Statist. Assoc.}}
  \bibinfo{volume}{84}, \bibinfo{number}{406} (\bibinfo{year}{1989}),
  \bibinfo{pages}{414--420}.
\newblock
\urldef\tempurl%
\url{https://doi.org/10.1080/01621459.1989.10478785}
\showDOI{\tempurl}


\bibitem[\protect\citeauthoryear{Kamousi, Lazard, Maheshwari, and
  Wuhrer}{Kamousi et~al\mbox{.}}{2016}]%
        {KAMOUSI20161}
\bibfield{author}{\bibinfo{person}{Pegah Kamousi}, \bibinfo{person}{Sylvain
  Lazard}, \bibinfo{person}{Anil Maheshwari}, {and} \bibinfo{person}{Stefanie
  Wuhrer}.} \bibinfo{year}{2016}\natexlab{}.
\newblock \showarticletitle{Analysis of farthest point sampling for
  approximating geodesics in a graph}.
\newblock \bibinfo{journal}{\emph{Computational Geometry}}
  \bibinfo{volume}{57} (\bibinfo{year}{2016}), \bibinfo{pages}{1--7}.
\newblock
\showISSN{0925-7721}
\urldef\tempurl%
\url{https://doi.org/10.1016/j.comgeo.2016.05.005}
\showDOI{\tempurl}


\bibitem[\protect\citeauthoryear{Kim and Lee}{Kim and Lee}{2010}]%
        {HARRA}
\bibfield{author}{\bibinfo{person}{Hung-sik Kim} {and} \bibinfo{person}{Dongwon
  Lee}.} \bibinfo{year}{2010}\natexlab{}.
\newblock \showarticletitle{\MakeUppercase{HARRA}: fast iterative hashed record
  linkage for large-scale data collections}. \bibinfo{pages}{525--536}.
\newblock
\urldef\tempurl%
\url{https://doi.org/10.1145/1739041.1739104}
\showDOI{\tempurl}


\bibitem[\protect\citeauthoryear{Kruskal}{Kruskal}{1964}]%
        {Kruskal1964}
\bibfield{author}{\bibinfo{person}{Joseph Kruskal}.}
  \bibinfo{year}{1964}\natexlab{}.
\newblock \showarticletitle{Multidimensional scaling by optimizing goodness of
  fit to a nonmetric hypothesis}.
\newblock \bibinfo{journal}{\emph{Psychometrika}} \bibinfo{volume}{29},
  \bibinfo{number}{1} (\bibinfo{date}{01 Mar} \bibinfo{year}{1964}),
  \bibinfo{pages}{1--27}.
\newblock
\showISSN{1860-0980}


\bibitem[\protect\citeauthoryear{Kumar, Zhang, and Nayar}{Kumar
  et~al\mbox{.}}{2008}]%
        {Kumar}
\bibfield{author}{\bibinfo{person}{Neeraj Kumar}, \bibinfo{person}{li Zhang},
  {and} \bibinfo{person}{Shree Nayar}.} \bibinfo{year}{2008}\natexlab{}.
\newblock \showarticletitle{What is a good nearest neighbors algorithm for
  finding similar patches in images?} \bibinfo{pages}{364--378}.
\newblock
\showISBNx{978-3-540-88685-3}
\urldef\tempurl%
\url{https://doi.org/10.1007/978-3-540-88688-4_27}
\showDOI{\tempurl}


\bibitem[\protect\citeauthoryear{Li, Jin, and Mehrotra}{Li
  et~al\mbox{.}}{2006}]%
        {Li-chen}
\bibfield{author}{\bibinfo{person}{Chen Li}, \bibinfo{person}{Liang Jin}, {and}
  \bibinfo{person}{Sharad Mehrotra}.} \bibinfo{year}{2006}\natexlab{}.
\newblock \showarticletitle{Supporting efficient record linkage for large data
  sets using mapping techniques}.
\newblock \bibinfo{journal}{\emph{World Wide Web}}  \bibinfo{volume}{9}
  (\bibinfo{date}{12} \bibinfo{year}{2006}), \bibinfo{pages}{557--584}.
\newblock
\urldef\tempurl%
\url{https://doi.org/10.1007/s11280-006-0226-8}
\showDOI{\tempurl}


\bibitem[\protect\citeauthoryear{Liang, Wang, Christen, and Gayler}{Liang
  et~al\mbox{.}}{2014}]%
        {Liang}
\bibfield{author}{\bibinfo{person}{Huizhi Liang}, \bibinfo{person}{Yanzhe
  Wang}, \bibinfo{person}{Peter Christen}, {and} \bibinfo{person}{Ross
  Gayler}.} \bibinfo{year}{2014}\natexlab{}.
\newblock \showarticletitle{Noise-tolerant approximate blocking for dynamic
  real-time entity resolution}. In \bibinfo{booktitle}{\emph{Advances in
  Knowledge Discovery and Data Mining}},
  \bibfield{editor}{\bibinfo{person}{Vincent~S. Tseng}, \bibinfo{person}{Tu~Bao
  Ho}, \bibinfo{person}{Zhi-Hua Zhou}, \bibinfo{person}{Arbee L.~P. Chen},
  {and} \bibinfo{person}{Hung-Yu Kao}} (Eds.). \bibinfo{publisher}{Springer
  International Publishing}, \bibinfo{address}{Cham},
  \bibinfo{pages}{449--460}.
\newblock


\bibitem[\protect\citeauthoryear{Loo}{Loo}{2014}]%
        {Loo2014}
\bibfield{author}{\bibinfo{person}{Mark P J Van~Der Loo}.}
  \bibinfo{year}{2014}\natexlab{}.
\newblock \showarticletitle{{The STRINGDIST package for approximate string
  matching}}.
\newblock \bibinfo{journal}{\emph{The R Journal}} \bibinfo{volume}{6},
  \bibinfo{number}{1} (\bibinfo{year}{2014}), \bibinfo{pages}{111--122}.
\newblock
\showISBNx{2073-4859}
\showISSN{20734859}


\bibitem[\protect\citeauthoryear{Mazeika and Böhlen}{Mazeika and
  Böhlen}{2006}]%
        {Mazeika}
\bibfield{author}{\bibinfo{person}{Arturas Mazeika} {and}
  \bibinfo{person}{Michael Böhlen}.} \bibinfo{year}{2006}\natexlab{}.
\newblock \showarticletitle{Cleansing databases of misspelled proper nouns}.
\newblock
\urldef\tempurl%
\url{https://doi.org/10.5167/uzh-56109}
\showDOI{\tempurl}


\bibitem[\protect\citeauthoryear{McCallum, Nigam, and Ungar}{McCallum
  et~al\mbox{.}}{2000}]%
        {MacCallum}
\bibfield{author}{\bibinfo{person}{Andrew McCallum}, \bibinfo{person}{Kamal
  Nigam}, {and} \bibinfo{person}{Lyle~H. Ungar}.}
  \bibinfo{year}{2000}\natexlab{}.
\newblock \showarticletitle{Efficient clustering of high-dimensional data sets
  with application to reference matching}. In
  \bibinfo{booktitle}{\emph{Proceedings of the Sixth ACM SIGKDD International
  Conference on Knowledge Discovery and Data Mining}} (Boston, Massachusetts,
  USA) \emph{(\bibinfo{series}{KDD '00})}. \bibinfo{publisher}{Association for
  Computing Machinery}, \bibinfo{address}{New York, NY, USA},
  \bibinfo{pages}{169–178}.
\newblock
\showISBNx{1581132336}
\urldef\tempurl%
\url{https://doi.org/10.1145/347090.347123}
\showDOI{\tempurl}


\bibitem[\protect\citeauthoryear{Papadakis, Skoutas, Thanos, and
  Palpanas}{Papadakis et~al\mbox{.}}{2020}]%
        {sur2}
\bibfield{author}{\bibinfo{person}{George Papadakis},
  \bibinfo{person}{Dimitrios Skoutas}, \bibinfo{person}{Emmanouil Thanos},
  {and} \bibinfo{person}{Themis Palpanas}.} \bibinfo{year}{2020}\natexlab{}.
\newblock \showarticletitle{Blocking and filtering techniques for entity
  resolution: A survey}.
\newblock \bibinfo{journal}{\emph{ACM Comput. Surv.}} \bibinfo{volume}{53},
  \bibinfo{number}{2}, Article \bibinfo{articleno}{31} (\bibinfo{date}{March}
  \bibinfo{year}{2020}), \bibinfo{numpages}{42}~pages.
\newblock
\showISSN{0360-0300}
\urldef\tempurl%
\url{https://doi.org/10.1145/3377455}
\showDOI{\tempurl}


\bibitem[\protect\citeauthoryear{Saeedi, Peukert, and Rahm}{Saeedi
  et~al\mbox{.}}{2017}]%
        {Saeedi}
\bibfield{author}{\bibinfo{person}{Alieh Saeedi}, \bibinfo{person}{Eric
  Peukert}, {and} \bibinfo{person}{Erhard Rahm}.}
  \bibinfo{year}{2017}\natexlab{}.
\newblock \showarticletitle{Comparative evaluation of distributed clustering
  schemes for multi-source entity resolution}.
\newblock
\showISBNx{978-3-319-66916-8}
\urldef\tempurl%
\url{https://doi.org/10.1007/978-3-319-66917-5_19}
\showDOI{\tempurl}


\bibitem[\protect\citeauthoryear{Silva and Tenenbaum}{Silva and
  Tenenbaum}{2002}]%
        {Silva_2002}
\bibfield{author}{\bibinfo{person}{Vin~de Silva} {and}
  \bibinfo{person}{Joshua~B. Tenenbaum}.} \bibinfo{year}{2002}\natexlab{}.
\newblock \showarticletitle{Global versus local methods in nonlinear
  dimensionality reduction}. In \bibinfo{booktitle}{\emph{Proceedings of the
  15th International Conference on Neural Information Processing Systems}}
  \emph{(\bibinfo{series}{NIPS'02})}. \bibinfo{publisher}{MIT Press},
  \bibinfo{address}{Cambridge, MA, USA}, \bibinfo{pages}{721–728}.
\newblock


\bibitem[\protect\citeauthoryear{Simonini, Papadakis, Palpanas, and
  Bergamaschi}{Simonini et~al\mbox{.}}{2018}]%
        {schema_agnostic}
\bibfield{author}{\bibinfo{person}{Giovanni Simonini}, \bibinfo{person}{George
  Papadakis}, \bibinfo{person}{Themis Palpanas}, {and} \bibinfo{person}{Sonia
  Bergamaschi}.} \bibinfo{year}{2018}\natexlab{}.
\newblock \showarticletitle{Schema-agnostic progressive entity resolution}.
\newblock \bibinfo{journal}{\emph{IEEE Transactions on Knowledge and Data
  Engineering}} \bibinfo{volume}{31}, \bibinfo{number}{6}
  (\bibinfo{year}{2018}), \bibinfo{pages}{1208--1221}.
\newblock


\bibitem[\protect\citeauthoryear{Talbert and Fisher}{Talbert and
  Fisher}{2000}]%
        {Talbert}
\bibfield{author}{\bibinfo{person}{Douglas Talbert} {and} \bibinfo{person}{Doug
  Fisher}.} \bibinfo{year}{2000}\natexlab{}.
\newblock \showarticletitle{An empirical analysis of techniques for
  constructing and searching k-dimensional trees}. \bibinfo{pages}{26--33}.
\newblock
\urldef\tempurl%
\url{https://doi.org/10.1145/347090.347100}
\showDOI{\tempurl}


\bibitem[\protect\citeauthoryear{Tang and Crovella}{Tang and Crovella}{2003}]%
        {Virtual_L}
\bibfield{author}{\bibinfo{person}{Liying Tang} {and} \bibinfo{person}{Mark
  Crovella}.} \bibinfo{year}{2003}\natexlab{}.
\newblock \showarticletitle{Virtual landmarks for the Internet}. In
  \bibinfo{booktitle}{\emph{Proceedings of the 3rd ACM SIGCOMM Conference on
  Internet Measurement}} (Miami Beach, FL, USA) \emph{(\bibinfo{series}{IMC
  '03})}. \bibinfo{publisher}{Association for Computing Machinery},
  \bibinfo{address}{New York, NY, USA}, \bibinfo{pages}{143–152}.
\newblock
\showISBNx{1581137737}
\urldef\tempurl%
\url{https://doi.org/10.1145/948205.948223}
\showDOI{\tempurl}


\bibitem[\protect\citeauthoryear{Wang, Chen, and Atabakhsh}{Wang
  et~al\mbox{.}}{2004}]%
        {cr_1}
\bibfield{author}{\bibinfo{person}{Gang Wang}, \bibinfo{person}{Hsinchun Chen},
  {and} \bibinfo{person}{Homa Atabakhsh}.} \bibinfo{year}{2004}\natexlab{}.
\newblock \showarticletitle{Automatically detecting deceptive criminal
  identities}.
\newblock \bibinfo{journal}{\emph{Commun. ACM}} \bibinfo{volume}{47},
  \bibinfo{number}{3} (\bibinfo{date}{March} \bibinfo{year}{2004}),
  \bibinfo{pages}{70–76}.
\newblock
\showISSN{0001-0782}
\urldef\tempurl%
\url{https://doi.org/10.1145/971617.971618}
\showDOI{\tempurl}


\end{thebibliography}


\end{document}